\documentclass[preprint2]{aastex}
\usepackage{spr-astr-addons}
\usepackage{url}
\bibpunct{(}{)}{,}{n}{,}{,}

\RequirePackage{color}

\begin{document}

 \title{Large retrograde Centaurs: visitors from the Oort cloud? 
        }

 \shorttitle{Large retrograde Centaurs}
 \shortauthors{de la Fuente Marcos and de la Fuente Marcos}

 \author{C.~de~la~Fuente Marcos} 
  \and 
 \author{R.~de~la~Fuente Marcos} 
 \affil{Universidad Complutense de Madrid, Madrid, Spain} 
 \email{nbplanet@fis.ucm.es}

 \begin{abstract}
    Among all the asteroid dynamical groups, Centaurs have 
    the highest fraction of objects moving in retrograde 
    orbits. The distribution in absolute magnitude, $H$, of 
    known retrograde Centaurs with semi-major axes in the 
    range 6--34 AU exhibits a remarkable trend: 10\% have 
    $H <$ 10 mag, the rest have $H >$ 12 mag. The largest 
    objects, namely (342842) 2008~YB$_{3}$, 2011~MM$_{4}$ 
    and 2013~LU$_{28}$, move in almost polar, very eccentric 
    paths; their nodal points are currently located near 
    perihelion and aphelion. In the group of retrograde 
    Centaurs, they are obvious outliers both in terms of 
    dynamics and size. Here, we show that these objects are 
    also trapped in retrograde resonances that make them 
    unstable. Asteroid 2013~LU$_{28}$, the largest, is a 
    candidate transient co-orbital to Uranus and it may be a 
    recent visitor from the trans-Neptunian region. 
    Asteroids 342842 and 2011~MM$_{4}$ are temporarily 
    submitted to various high-order retrograde resonances 
    with the Jovian planets but 342842 may be ejected 
    towards the trans-Neptunian region within the next few 
    hundred kyr. Asteroid 2011~MM$_{4}$ is far more stable. 
    Our analysis shows that the large retrograde Centaurs 
    form an heterogeneous group that may include objects 
    from various sources. Asteroid 2011~MM$_{4}$ could be a 
    visitor from the Oort cloud but an origin in a 
    relatively stable closer reservoir cannot be ruled out. 
    Minor bodies like 2011~MM$_{4}$ may represent the 
    remnants of the primordial planetesimals and signal the 
    size threshold for catastrophic collisions in the early 
    Solar System.  
 \end{abstract}

 \keywords{Celestial mechanics $\cdot$ 
           Minor planets, asteroids: general $\cdot$
           Minor planets, asteroids: individual: (342842)~2008~YB$_{3}$ $\cdot$
           Minor planets, asteroids: individual: 2011~MM$_{4}$ $\cdot$
           Minor planets, asteroids: individual: 2013~LU$_{28}$ $\cdot$ 
           Planets and satellites: individual: Uranus
          }

 \section{Introduction}
    In our planetary system, most objects go around the Sun following counterclockwise (prograde) orbits as viewed from above the north 
    pole of our star. However, a large number of comets orbit the Sun clockwise (retrograde) and a substantial percentage of these objects 
    have highly inclined orbits. They are believed to have their origin in the Oort cloud (see e.g. Duncan 2008), a vast and remote 
    spherical reservoir of cometary material tens of thousands of astronomical units in diameter that completely surrounds the Solar System. 

    Out of 3265 objects currently classified by the Jet Propulsion Laboratory (JPL) Small-Body Database\footnote{http://ssd.jpl.nasa.gov/sbdb.cgi} 
    as comets, 1918 move in retrograde orbits (58.7\%). For long-period comets the retrograde fraction is 49.7\%, for Halley-type comets
    is 35.7\% and for Jupiter-family comets is 16.7\%. In contrast, only 53 of the 641502 currently catalogued asteroids orbit the Sun 
    clockwise (0.008\%). For asteroids in the outer Solar System (catalogued as Centaurs or trans-Neptunian objects) the fraction of 
    retrograde objects is 2.7\% (48 out of 1804). Therefore, most retrograde asteroids have been found in the region of the giant planets 
    and beyond. At 11.0\% (30 out of 272 catalogued objects in the JPL Small-Body Database), Centaurs have the highest retrograde fraction 
    among all the dynamical groups or asteroid families.

    Centaurs are a transient asteroidal population whose orbits cross those of the outer planets (see e.g. Di Sisto and Brunini 2007). Most 
    Centaurs may have their origin in the trans-Neptunian belt, some may have come from the Oort cloud. The main source of Centaurs remains 
    to be discovered. They are probably an heterogeneous population with multiple sources (see e.g. Di Sisto et al. 2010), perhaps none 
    being dominant (but see Horner and Lykawka 2010). 

    The most intriguing of the known Centaurs are those moving in retrograde orbits. The study of retrograde Centaurs, especially those 
    following high inclination trajectories, may represent a rare opportunity to look at objects that have suffered little surface 
    alteration over the age of the Solar System (Sheppard 2010). The three largest known (in terms of absolute magnitude, $H$) retrograde 
    Centaurs are (342842) 2008~YB$_{3}$ ($H$ = 9.3 mag), 2011~MM$_{4}$ ($H$ = 9.3 mag) and 2013~LU$_{28}$ ($H$ = 8.1 mag); they all move in 
    near-polar orbits. Minor bodies following eccentric, high-inclination paths spend most of the time well beyond the plane of the Solar 
    System and away from the destabilizing influence of the giant planets, although close encounters with these planets are still possible 
    at the nodes.

    Here, we show that, far from being just dynamical curiosities, the three largest retrograde Centaurs may be the key to identify the size 
    threshold for catastrophic collisions in the early Solar System. On the other hand, they are also submitted to retrograde mean motion 
    resonances that make them dynamically unstable, emphasizing the fact that these objects have not remained in their current orbits for 
    long. This paper is organized as follows. In Sect. 2, we present a comparative statistical analysis of the properties of known 
    prograde and retrograde Centaurs that reveals compelling reasons to single out the largest retrograde objects. The current dynamical 
    status of these large retrograde Centaurs is studied in Sect. 3, where the numerical model used in our $N$-body simulations is also 
    briefly discussed. In Sect. 4, we consider the implications of our results and in Sect. 5 we summarize our conclusions.

 \section{Retrograde Centaurs}
\begin{figure*}
      \centering
       \includegraphics[width=0.49\linewidth]{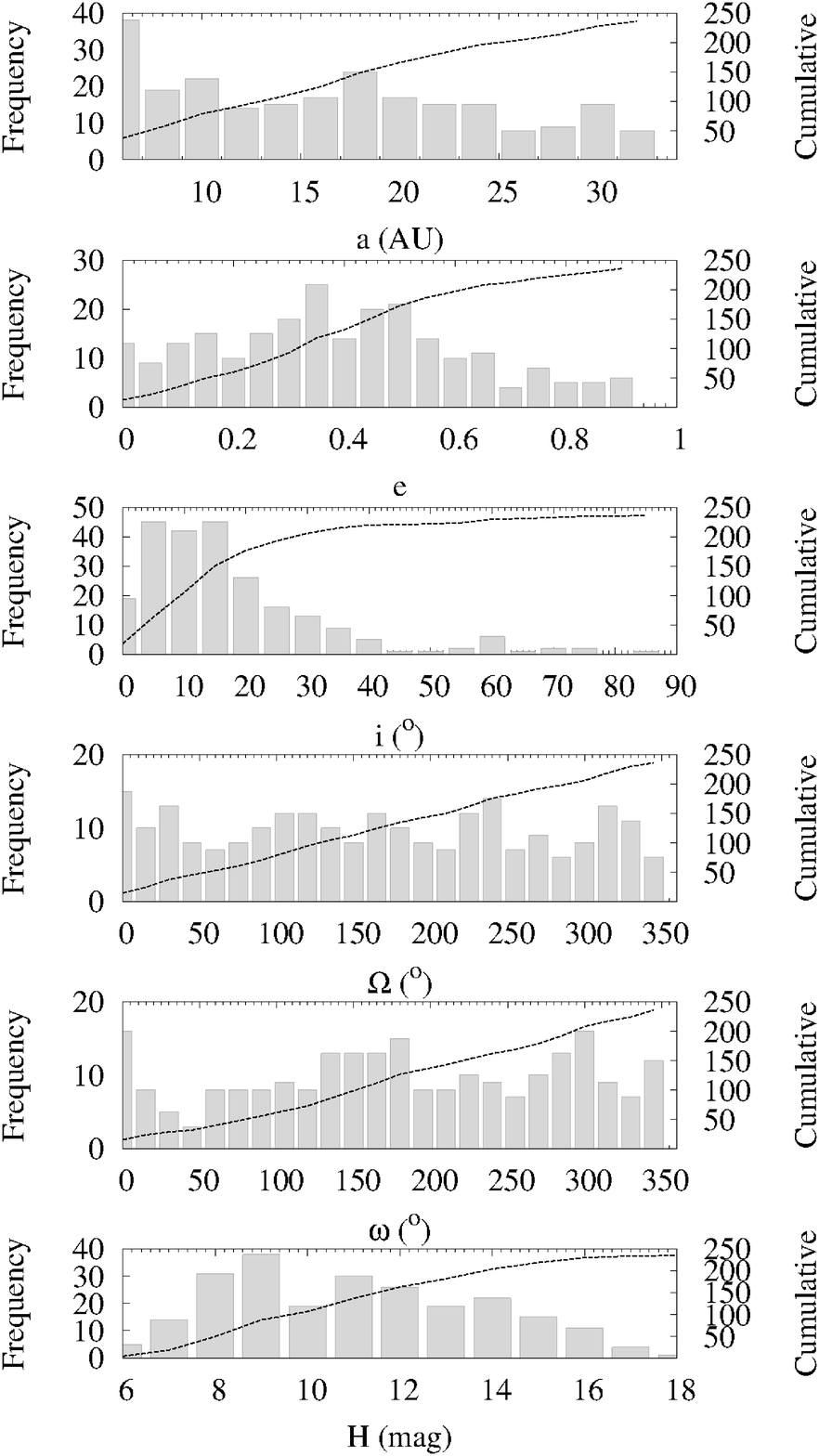}
       \includegraphics[width=0.49\linewidth]{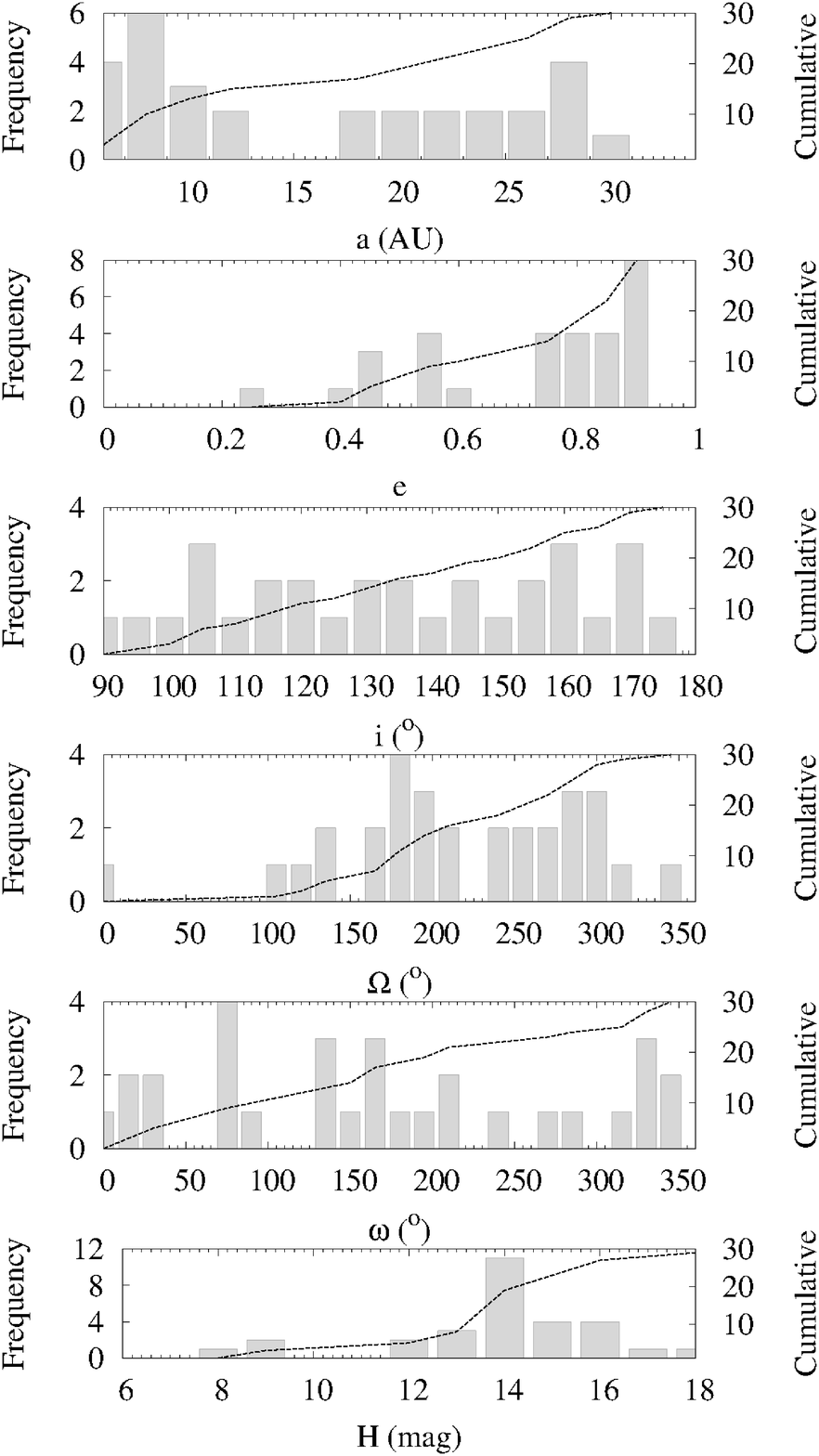}
       \caption{Distributions in semi-major axis, $a$, eccentricity, $e$, inclination, $i$, longitude of the ascending node, $\Omega$,  
                argument of perihelion, $\omega$, and absolute magnitude, $H$, of observed Centaurs following prograde (left-hand panels)
                and retrograde orbits (right-hand panels). 
               }
       \label{pro-ret}
    \end{figure*}

    Following Alexandersen et al. (2013), let us consider the group of minor bodies whose semi-major axes are between 6 and 34 AU. There are 
    266 such asteroids currently (as of 2014 May 5) in the JPL Small-Body Database. This number is slightly less than the total number of
    known Centaurs quoted above. Out of them, 236 (88.7\%) follow prograde orbits and 30 (11.3\%) follow retrograde orbits. Figure 
    \ref{pro-ret} shows the distribution in orbital parameter space and $H$ of prograde (left-hand panels) and retrograde (right-hand 
    panels) Centaurs. Even if the number of known retrograde Centaurs is small, it is highly improbable that the two Centaur populations 
    have a common origin. The distribution in absolute magnitude, which is a proxy for size, clearly shows that retrograde Centaurs are 
    unlikely to be the result of gravitational scattering of originally prograde Centaurs, unless we assume that smaller objects are more 
    efficiently scattered to become retrograde than larger ones (see the $e$- and $i$-panels) or that smaller retrograde objects are easier 
    to identify than their prograde counterparts. 
%
%
%
%

    Most prograde orbits have eccentricities in the range 0.2--0.6 and relatively low inclinations, $<$20\degr. The small but noticeable 
    group of prograde orbits with inclination $\sim$63\degr in Fig. \ref{pro-ret} (left-hand panel) is consistent with predictions made by 
    e.g. Gallardo (2006) on the existence of objects submitted to the Kozai resonance at that critical inclination. The relative excess of 
    objects with argument of perihelion close to 0\degr or 180\degr could be due to observational bias because only objects with 
    declinations $|\delta|<$24\degr are observed (see de la Fuente Marcos and de la Fuente Marcos 2014). 

    The raw distribution in $H$ of prograde orbits shows evidence of the so called divot or sudden decrease in numbers for $H\sim$ 9 mag (or 
    a diameter of nearly 100 km) as found by e.g. Shankman et al. (2013). This has been interpreted by those authors as the relic signature 
    of the size distribution of the primordial planetesimal population, with the larger objects being the remnants. A divot in the size 
    distribution of trans-Neptunian objects was predicted by Fraser (2009) using the results of collisional evolution calculations. A lack 
    of objects in the size range 1--20 km was also anticipated by Charnoz and Morbidelli (2007) for minor bodies in the scattered disk and 
    the Oort cloud. This feature had been previously identified and confirmed observationally for the size range 25--60 km by Bernstein et 
    al. (2004), Fuentes and Holman (2008), and Fraser and Kavelaars (2009). This break in the size distribution may signal the onset of 
    collisional evolution. Schlichting et al. (2013) consider that, for trans-Neptunian objects, the observed size distribution above a 
    radius of $\sim$30 km is primordial. They also found that there is a significant deficit of bodies with size $\sim$10 km and a strong 
    excess of bodies of nearly 2 km in radius; this excess leaves a permanent signature in the size distribution of these objects that is 
    still clearly observable after 4.5 Gyr of collisional evolution. Fraser et al. (2014) have found that the observed size distributions 
    are difficult to explain within the currently accepted standard model of planetesimal formation. 
  
    The distribution in $H$ for retrograde orbits in Fig. \ref{pro-ret} (left-hand panel) exhibits a most unusual behaviour. There are only 
    three objects (10\%) with $H <$ 10 mag: (342842) 2008~YB$_{3}$ ($H$ = 9.3 mag), 2011~MM$_{4}$ ($H$ = 9.3 mag) and 2013~LU$_{28}$ ($H$ = 
    8.1 mag). The rest have $H >$ 12 mag (73.3\% have $H >$ 14 mag). Although the number of known retrograde Centaurs is small, this result 
    can be seen as a dramatic confirmation of the existence of a break in the size distribution as predicted by e.g. Fraser (2009). There is 
    an obvious deficit of objects with $H$ in the range 10--14 mag although they must be significantly easier to detect, if they do indeed 
    exist, than fainter objects of a size below 10 km. Although the available sample is far from complete and likely biased, it should not 
    be biased in favour of 1-km objects against those in the size range 10--100 km. The scarcity of intermediate-size objects must be real 
    and it may have a collisional, dynamical or even primordial origin (or a combination of the three). In principle, and following Fraser 
    (2009), Schlichting et al. (2013) or Shankman et al. (2013), some or all the large retrograde Centaurs may represent a remnant 
    population of primordial planetesimals. 

\begin{figure}
      \centering
       \includegraphics[width=\linewidth]{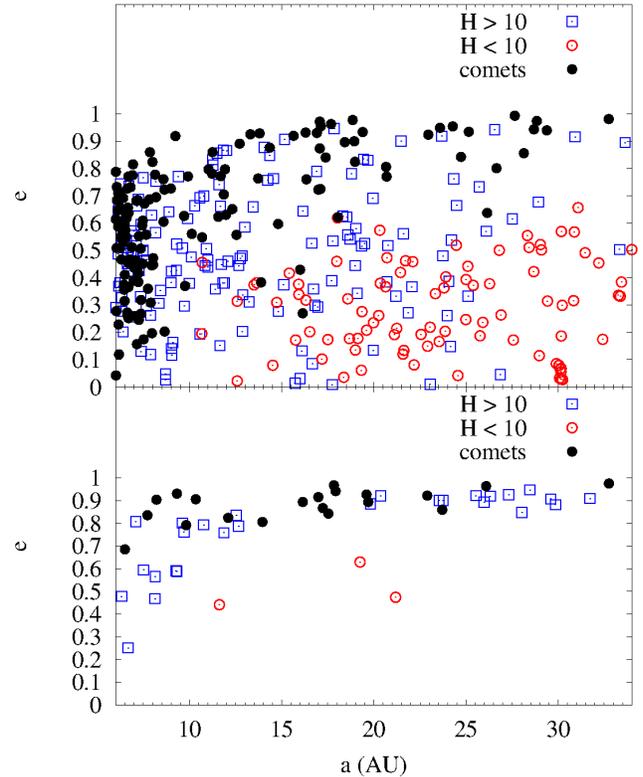}
       \caption{The distribution in the ($a$, $e$) plane of prograde (top panel) and retrograde (bottom panel) Centaurs and comets with 
                semi-major axes in the range 6--34 AU. There are 145 prograde and 20 retrograde comets in that semi-major axis range.
                See also Fig. \ref{real}. The three largest retrograde Centaurs, namely (342842)~2008~YB$_{3}$, 2011~MM$_{4}$ and 
                2013~LU$_{28}$, are clear dynamical outliers; they have the largest perihelia.
               }
       \label{aeCEN}
    \end{figure}

\begin{figure}
        \centering
         \includegraphics[width=\linewidth]{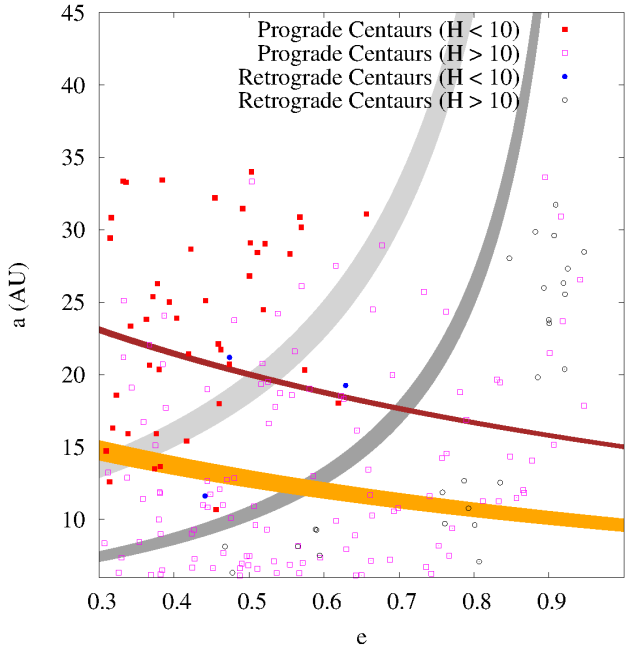}
         \caption{Positions of the group of present-day Centaurs discussed in Sect. 2 in the ($e$, $a$) plane. The dark gray area represents 
                  the eccentricity/semi-major axis combination with periapsis between the perihelion and aphelion of Jupiter, the light gray 
                  area shows the equivalent parameter domain if Saturn is considered instead of Jupiter. The orange area corresponds to the 
                  ($e$, $a$) combination with apoapsis between the perihelion and aphelion of Uranus and the brown area shows the 
                  counterpart for Neptune.
                 }
         \label{real}
      \end{figure}

    In addition, the distribution in $H$ of retrograde Centaurs suggests that at least two distinct populations of rather different 
    dynamical origin are present. In support of this interpretation, Fig. \ref{aeCEN} (see also Fig. \ref{real}) shows the distribution in 
    the ($a$, $e$) plane of both groups and comets in the same semi-major axis range: the three largest retrograde Centaurs clearly stand 
    out (bottom panel). The brightest (largest) objects also appear to be dynamically different. The distribution of semi-major axes and 
    eccentricities of retrograde objects with $H >$ 12 resembles that of members of the detached or scattered disk populations (Levison and 
    Duncan 1997; Gomes 2011). In terms of the size of their members, these trans-Neptunian populations appear to be of collisional origin 
    and, dynamically, they may have reached their current state after suffering strong gravitational perturbations. The surprising abundance 
    of retrograde minor bodies at $H\sim14$ mag may signal a high density of objects with size $\sim1$ km as predicted by Fraser (2009) or, 
    more recently, Belton (2014).
%
%
%
%
%
%
%
%
   
    It may be argued that we unfoundedly invoked the observed decrease in the number of prograde Centaurs at magnitude $H\sim$ 9 to 
    implicitly justify the division into two groups ($H <$ 10 mag and $H >$ 10 mag) in Fig. \ref{aeCEN}. However, such concern is completely
    unjustified as Fig. \ref{aeCEN} clearly shows. It is rather obvious from the figure that the three largest retrograde Centaurs are clear 
    dynamical outliers because they have the largest perihelia. Out of the three largest retrograde Centaurs, only 342842 is a relatively 
    well-studied object (Sheppard 2010; Pinilla-Alonso et al. 2013; Bauer et al. 2013). But, what is special or unique about these objects? 
    Both 342842 and 2013~LU$_{28}$ can suffer close encounters with Jupiter; 2011~MM$_{4}$ may be considerably more stable.

    Brasser et al. (2012) and Volk and Malhotra (2013) have found that retrograde or very high inclination Centaurs cannot have evolved from 
    the trans-Neptunian belt due to close encounters with the Jovian planets. Instead, they favour a source in the Oort cloud for these 
    objects. In the Solar System, near-polar orbits are almost dynamically frozen in time and, potentially, they could be stable for 
    hundreds of Myr. This is particularly true if the nodes of the orbits of the objects remain relatively far from the paths of the 
    planets. However, if the objects following these unusual trajectories are also submitted to mean motion resonances, their dynamical 
    stability could be compromised even if they do not undergo close encounters with planetary bodies because they may experience chaotic 
    diffusion. 

    Morais and Namouni (2013a) have brought to the attention of the astronomical community that minor bodies following retrograde orbits 
    around the Sun may also be trapped in a mean motion resonance with a planet. The same authors have identified several asteroids 
    currently moving in retrograde resonance with Jupiter and Saturn (Morais and Namouni 2013b). This discovery is of considerable practical 
    interest because of its potential impact on our understanding of the early stages of the evolution of the Solar System and also on the 
    dynamics of transient populations like the Centaurs. Are the largest retrograde Centaurs perhaps submitted to retrograde mean motion 
    resonances with the giant planets?

 \section{Dynamical evolution of large resonant retrograde Centaurs}
    The dynamics of resonant retrograde minor bodies has been studied by Morais and Namouni (2013a, 2013b). For an object in retrograde 
    resonance with a planet that moves in a prograde orbit, the resonant argument can be written as $\sigma = q \lambda - p \lambda_{\rm P} 
    - (q + p - 2 k) \varpi + 2 k \Omega$, where $\lambda$ and $\lambda_{\rm P}$ are the mean longitudes of the retrograde object and the 
    prograde planet, respectively, $\varpi$ is the longitude of perihelion of the retrograde object, $\varpi = \omega - \Omega$, $\Omega$ 
    is the longitude of the ascending node and $\omega$ is the argument of perihelion. The mean longitude of a prograde object is given by 
    $M + \Omega + \omega$, where $M$ is the mean anomaly, but the one for a retrograde object is written as $M + \omega - \Omega$. The 
    numbers $p$, $q$ and $k$ are integers with $p + q \geq 2 k$. If the two objects are in resonance, $T/T_{\rm P} = p/q$, where $T$ and 
    $T_{\rm P}$ are the orbital periods of the retrograde object and the prograde planet, respectively. If, over a given period of time, 
    the resonant argument, $\sigma$, can take any value in the range 0--360\degr, then it circulates but if it exhibits oscillations 
    (symmetric or asymmetric) around a certain value (usually 0\degr or 180\degr), then we say that it librates. If the resonant argument 
    librates, then the object is trapped in a $p$:$q$ retrograde (or $p$:$-q$) mean motion resonance with the planet. Retrograde resonances 
    are weaker than their prograde counterparts (Morais and Namouni 2013a, 2013b).
%
%
     \begin{table*}
       \centering
        \fontsize{8}{11pt}\selectfont
        \tabcolsep 0.35truecm
        \caption{Heliocentric ecliptic Keplerian orbital elements of large retrograde Centaurs (342842)~2008~YB$_{3}$, 2011~MM$_{4}$ and 
                 2013~LU$_{28}$. Values include the 1$\sigma$ uncertainty (Epoch = JD2456800.5, 2014-May-23.0; J2000.0 ecliptic and 
                 equinox. Data for 2013~LU$_{28}$ are referred to epoch 2456455.5, 2013-Jun-12.0. Source: JPL Small-Body Database.)
                }
        \begin{tabular}{lllll}
         \hline
                                                            &   & (342842) 2008~YB$_{3}$  &   2011~MM$_{4}$     &   2013~LU$_{28}$ \\
         \hline
          Semi-major axis, $a$ (AU)                         & = &  11.61769$\pm$0.00007   &  21.183$\pm$0.011   &  19$\pm$18       \\
          Eccentricity, $e$                                 & = &   0.440962$\pm$0.000003 &   0.4739$\pm$0.0004 &   0.6$\pm$0.5    \\
          Inclination, $i$ (\degr)                          & = & 105.02980$\pm$0.00002   & 100.4457$\pm$0.0002 & 117$\pm$6        \\
          Longitude of the ascending node, $\Omega$ (\degr) & = & 112.498683$\pm$0.000010 & 282.6016$\pm$0.0005 & 266$\pm$7        \\
          Argument of perihelion, $\omega$ (\degr)          & = & 330.70350$\pm$0.00009   &   7.07$\pm$0.02     & 179$\pm$38       \\
          Mean anomaly, $M$ (\degr)                         & = &  29.3043$\pm$0.0003     &  34.75$\pm$0.03     & 298$\pm$142      \\
          Perihelion, $q$ (AU)                              & = &   6.494734$\pm$0.000003 &  11.144$\pm$0.004   &   7$\pm$4        \\
          Aphelion, $Q$ (AU)                                & = &  16.74064$\pm$0.00011   &  31.22$\pm$0.02     &  31$\pm$30       \\
          Absolute magnitude, $H$ (mag)                     & = &   9.3                   &   9.3               &   8.1            \\
         \hline
        \end{tabular}
        \label{elements}
     \end{table*}
%
%

    In this work, the orbital evolution of the three largest retrograde Centaurs is computed using the Hermite integration scheme described 
    by Makino (1991) and implemented by Aarseth (2003). The standard version of this $N$-body sequential code is publicly available from the 
    IoA web site\footnote{http://www.ast.cam.ac.uk/$\sim$sverre/web/pages/nbody.htm}. Our numerical simulations include the gravitational 
    perturbations of the eight major planets, the Moon, the barycentre of the Pluto-Charon system and the three largest asteroids (dwarf 
    planet (1) Ceres, (2) Pallas and (4) Vesta). Initial positions and velocities are based on the DE405 planetary orbital ephemerides 
    (Standish 1998) referred to the barycentre of the Solar System. Together with the orbital elements, they have been obtained from the JPL 
    on-line Solar System data service\footnote{http://ssd.jpl.nasa.gov/?planet\_pos} (Giorgini et al. 1996). In the integrations, the 
    relative error in the total energy was always as low as $3.0\times10^{-15}$ or lower after a simulated time of 1 Myr. The corresponding 
    error in the total angular momentum is several orders of magnitude smaller. Further details on the physical model and the validation of 
    the code can be found in de la Fuente Marcos and de la Fuente Marcos (2012a). Besides the calculations performed using the nominal 
    orbits as provided by the JPL Solar System data service, we have followed the evolution of 50 additional control orbits obtained from 
    the nominal ones and the quoted uncertainties (see Table \ref{elements}). Their orbital elements have been sampled from a 
    six-dimensional Gaussian distribution around the nominal ones (within 3$\sigma$ with the exception of 2013~LU$_{28}$). In all the 
    figures, $t$ = 0 coincides with the JD2456800.5 epoch unless explicitly stated. The dynamical evolution of the objects is followed for 
    $\pm$500 kyr.
%
%
      \begin{figure}
        \centering
         \includegraphics[width=\linewidth]{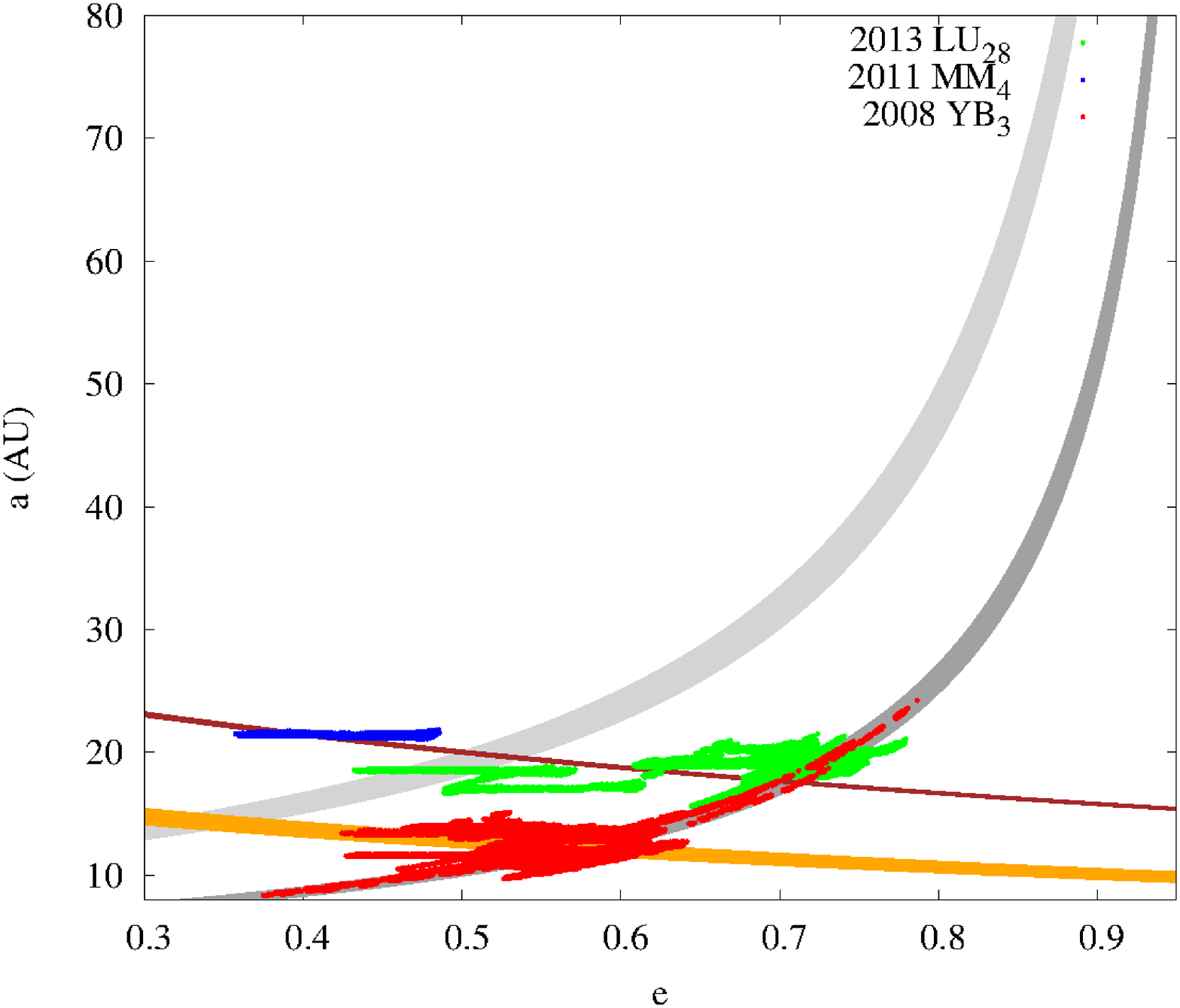}
         \caption{Orbital evolution of (342842) 2008~YB$_{3}$, 2011~MM$_{4}$ and 2013~LU$_{28}$ in the time interval (-500, 500) kyr.
                  Relatively long episodes of horizontal (resonant) oscillations are visible. For 2011~MM$_{4}$, no large changes in the 
                  orbital elements are observed over the studied time interval. The dark gray area represents the ($e$, $a$) combination 
                  with periapsis between the perihelion and aphelion of Jupiter, the light gray area shows the equivalent parameter domain 
                  if Saturn is considered instead of Jupiter. The orange area corresponds to the ($e$, $a$) combination with apoapsis 
                  between the perihelion and aphelion of Uranus and the brown area shows that of Neptune. These results correspond to the 
                  nominal orbits in Table \ref{elements}. 
                 }
         \label{ea}
      \end{figure}
%
%
%
%
      \begin{figure}
        \centering
         \includegraphics[width=\linewidth]{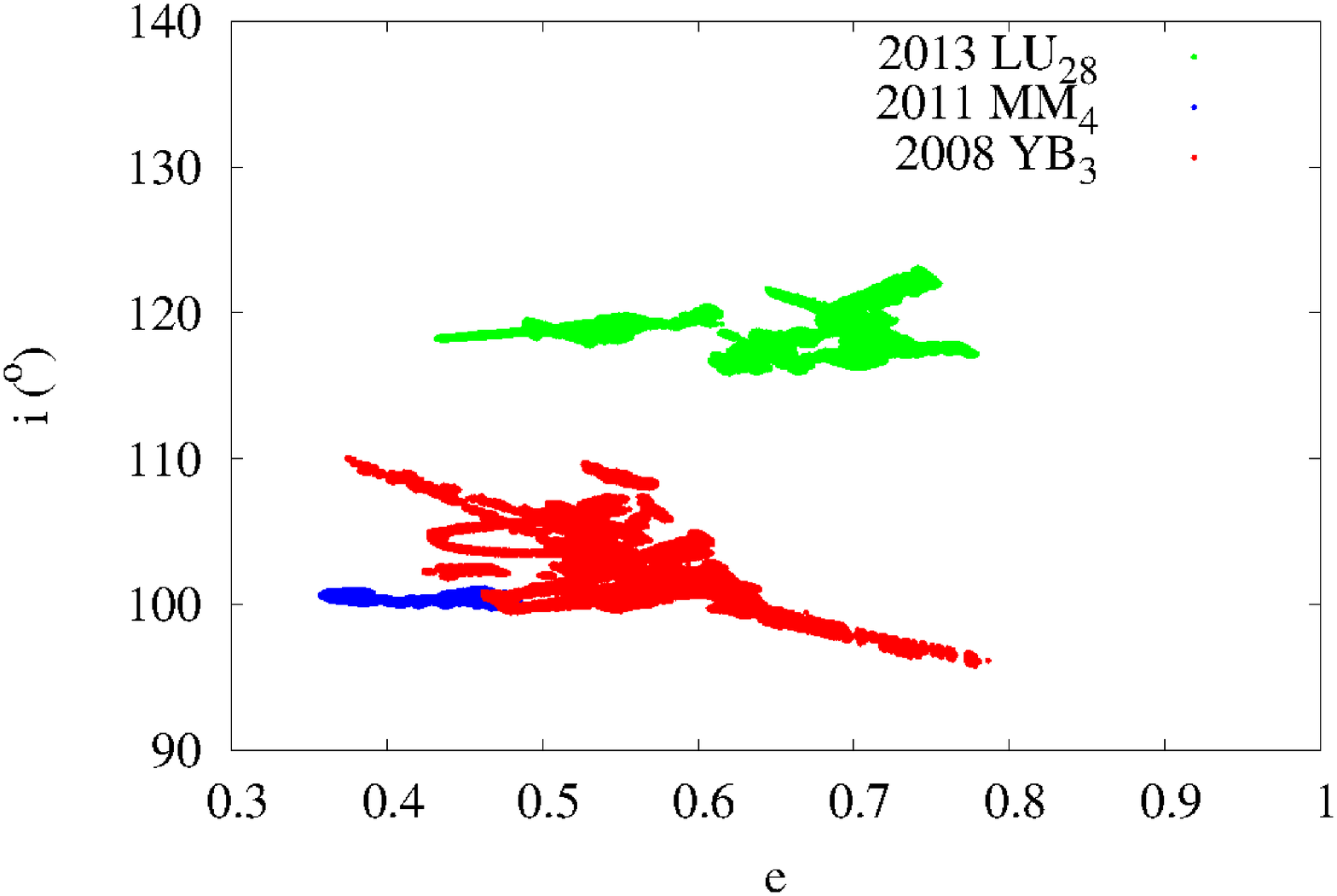}
         \caption{Similar to Fig. \ref{ea} but for the ($e$, $i$) evolution.
                 }
         \label{ei}
      \end{figure}
%
%

\begin{figure*}
       \centering
        \includegraphics[width=\linewidth]{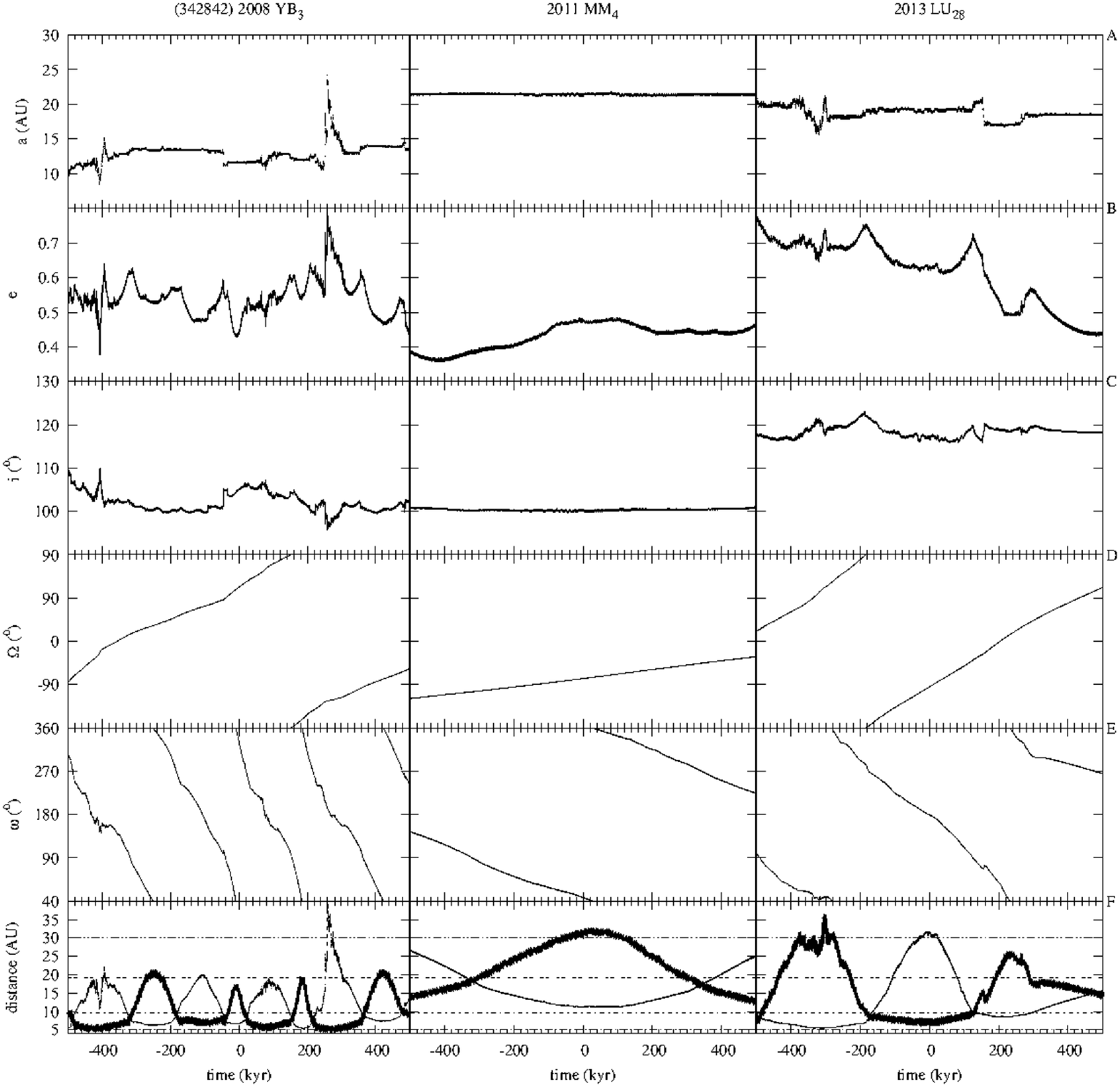}
        \caption{Comparative time evolution of various parameters for the three largest retrograde Centaurs: (342842)~2008~YB$_{3}$ (left), 
                 2011~MM$_{4}$ (centre) and 2013~LU$_{28}$ (right). The orbital elements $a$ (panel A), $e$ (panel B), $i$ (panel C), 
                 $\Omega$ (panel D) and $\omega$ (panel E). The distances to the ascending (thick line) and descending nodes (dotted line) 
                 appear in panel F. The semi-major axes of the giant planets are also shown. The output time-step for these plots is 2 yr. 
                 These results correspond to the nominal orbits in Table \ref{elements}.
                }
        \label{all}
     \end{figure*}

      Figures \ref{ea}, \ref{ei} and \ref{all} evidence that the overall orbital evolution of the three objects is rather different although 
      all of them experience horizontal oscillations in the ($e$, $a$) plane as secular resonances modify $e$ at constant $a$. These figures 
      also show that, when $\omega$ is close to 0\degr or 180\degr, (342842)~2008~YB$_{3}$ can experience close encounters with Jupiter at 
      perihelion and with Uranus at aphelion, 2011~MM$_{4}$ can encounter Saturn at perihelion and Neptune at aphelion, and 2013~LU$_{28}$ 
      can find Jupiter at perihelion and Neptune at aphelion (see also Fig. \ref{real}). Encounters can only take place at the nodes of the 
      orbit. For a retrograde orbit, the distance between the Sun and the nodes (Fig. \ref{all}, F-panels) is given by $r = a (1 - e^2) / (1 
      \pm e \cos \omega)$, where the `+' sign denotes the descending node and the `-' sign the ascending node. In the following, we explore 
      the dynamics of these unusual objects in more detail.
%
%
%
%

      \subsection{(342842) 2008~YB$_{3}$}
         Asteroid (342842) 2008~YB$_{3}$ was discovered on 2008 December 18 by R. H. McNaught observing for the Siding Spring Survey with
         the 0.5-m Uppsala Schmidt (McNaught et al. 2008). When first observed, it had a $V$ magnitude of 18.1. It was soon obvious that it 
         follows a near-polar retrograde orbit characterized by $a$ = 11.62 AU, $e$ = 0.44 and $i$ = 105\degr (see Table \ref{elements}). 
         Its orbit is well determined with 466 observations spanning a data-arc of 1856 d. It is relatively large with $H$ = 9.3 mag or a 
         diameter of 67 km for an albedo of 6.2\%, although it is probably larger. Sheppard (2010) found moderately red colours for this 
         minor body, consistent with those of the neutral group of the Centaurs, and concluded that due to its relatively short perihelion, 
         the object may have undergone surface sublimation and interior recrystallization. Pinilla-Alonso et al. (2013) used photometry, 
         spectroscopy and numerical simulations to show that this asteroid has a rotational period of 16.4 hr and, when compared with solar 
         colours, is moderately red with low albedo (perhaps depleted of ices), and it came from the trans-Neptunian region. 
         
\begin{figure}
        \centering
         \includegraphics[width=\linewidth]{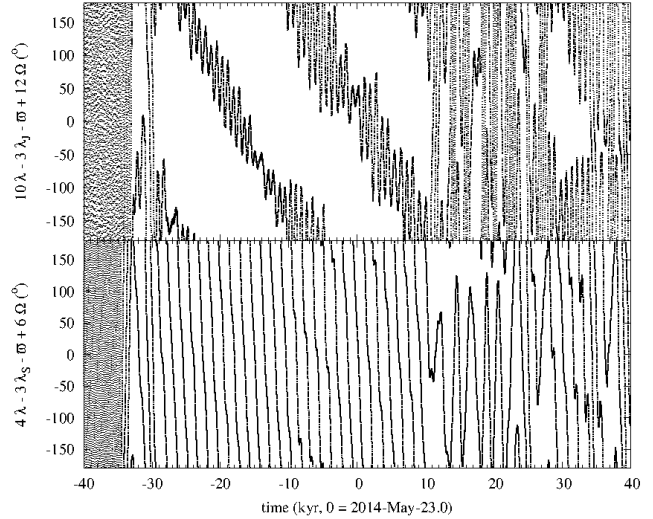}
         \caption{Evolution of resonant arguments for (342842) 2008~YB$_{3}$.
                 }
         \label{yb3}
      \end{figure}

         Asteroid 342842 is currently trapped in an accidental 3:10 retrograde resonance with Jupiter (libration around 0\degr, see Fig. 
         \ref{yb3}, top panel). We call it accidental because the resonant argument circulates with a superimposed short-period oscillation. 
         This behaviour is similar to the one found for Plutino (15810) 1994 JR$_{1}$, a quasi-satellite of Pluto (de la Fuente Marcos and
         de la Fuente Marcos 2012b). This resonant episode with Jupiter started nearly 30 kyr ago and it will cease in about 10 kyr. Just 
         after the episode, this minor body will be submitted to a 3:4 retrograde resonance with Saturn (asymmetric libration around 0\degr, 
         see Fig. \ref{yb3}, bottom panel) for nearly 10 kyr. Close encounters with Jupiter and more distant flybys with Saturn trigger the 
         ejection from the retrograde resonances with Jupiter and Saturn. All the control orbits exhibit similar behaviour during the 
         time-span plotted in Fig. \ref{yb3} but the dynamical evolution after the close encounters differs significantly between control 
         orbits. In about half the cases, the 3:4 retrograde resonance with Saturn is observed prior and after the 3:10 retrograde resonance 
         with Jupiter. The time-scale necessary for two initially infinitesimally close trajectories associated with this object to separate 
         appreciably, or $e$-folding time, is $\sim$3 kyr. The object will be ejected towards the trans-Neptunian region within the next 
         few hundred kyr. Our results indicate that this object is less dynamically stable than implied in Pinilla-Alonso et al. (2013).
%
%
%
%

      \subsection{2011~MM$_{4}$}
         Asteroid 2011~MM$_{4}$ was discovered on 2011 June 24 by N. Primak, A. Schultz, S. Watters, J. Thiel and T. Goggia observing for 
         the Pan-STARRS 1 project with the 1.8-m Ritchey-Chretien telescope from Haleakala (Woodworth et al. 2011). When discovered, its $r$ 
         magnitude was 21.1. It follows a path characterized by $a$ = 21.21 AU, $e$ = 0.47 and $i$ = 100\degr (see Table \ref{elements}). 
         Its orbit is relatively well determined with 92 observations spanning a data-arc of 396 d. With $H$ = 9.3 mag, it is very similar 
         in size to (342842) 2008~YB$_{3}$ although probably larger (Bauer et al. 2013). 
         
\begin{figure}
        \centering
         \includegraphics[width=\linewidth]{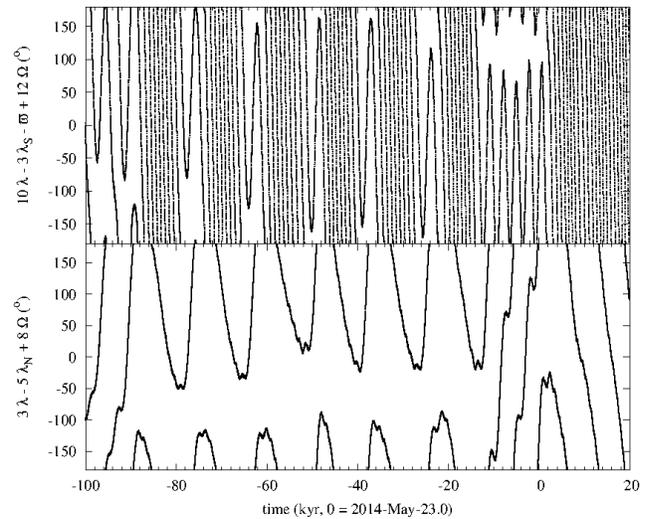}
         \caption{Evolution of resonant arguments for 2011~MM$_{4}$.
                 }
         \label{mm4}
      \end{figure}

         Asteroid 2011~MM$_{4}$ is currently trapped in the 3:10 retrograde resonance with Saturn (asymmetric libration around 90\degr, see 
         Fig. \ref{mm4}, top panel). This resonant episode started nearly 14 kyr ago and it will end in about 3 kyr. Prior to this episode,
         we observe that the critical angle alternates between circulation and asymmetric libration, indicating that the motion is chaotic.
         Before and after the resonant interaction with Saturn, this minor body was submitted to the 5:3 retrograde resonance with Neptune 
         (asymmetric libration around 180\degr, see Fig. \ref{mm4}, bottom panel). Relatively close encounters with Saturn and Neptune are 
         responsible for the ejection from the resonances. During the entire resonant phase, the argument of perihelion remains close to 
         0\degr, suggesting a behaviour similar to that of a Kozai secular resonance (Kozai 1962) although both $e$ and $i$ remain 
         relatively constant (see Fig. \ref{all}). As in the previous case, all the control orbits exhibit similar behaviour during the 
         time-span plotted in Fig. \ref{mm4}. In some cases, the 3:10 retrograde resonance with Saturn results in asymmetric libration 
         around 180\degr instead of 90\degr.

\begin{figure}
       \centering
        \includegraphics[width=\linewidth]{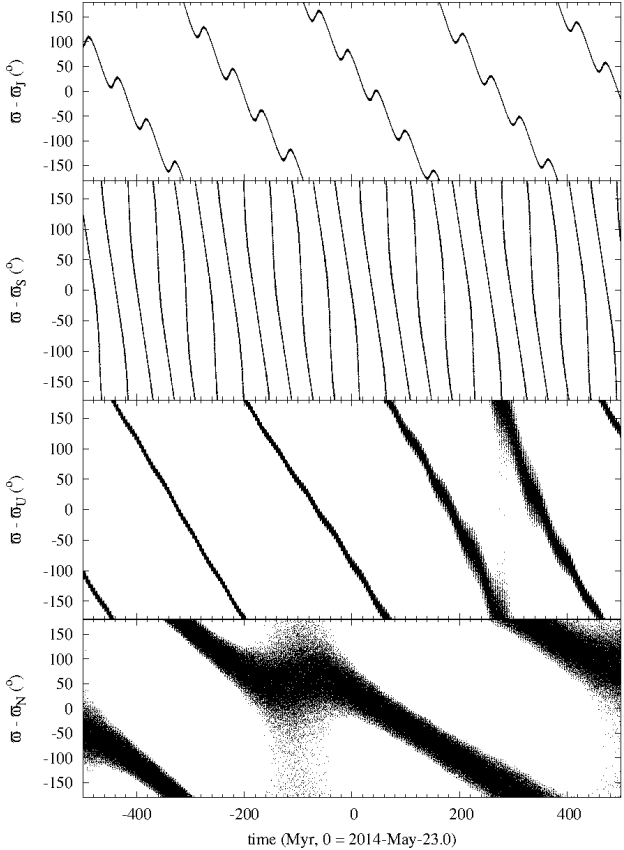}
        \caption{Time evolution of the relative longitude of the perihelion, $\Delta \varpi$, of 2011~MM$_{4}$ with respect to the giant 
                 planets: referred to Jupiter ($\varpi - \varpi_{J}$), to Saturn ($\varpi - \varpi_{S}$), to Uranus ($\varpi - \varpi_{U}$), 
                 and to Neptune ($\varpi - \varpi_{N}$). In all cases, the resonant argument $\Delta \varpi$ circulates over the entire 
                 simulated time interval although it is nearly in secular resonance with Jupiter. These results are for the nominal orbit in 
                 Table \ref{elements}.
                }
        \label{rellon}
     \end{figure}

         Our integrations indicate that this object can remain dynamically stable for several Myr although this issue will be studied in 
         more detail later. Its $e$-folding time is $\sim$10 kyr though. It is not submitted to any typical secular resonance (see Fig. 
         \ref{rellon}). The cause of the relative stability of this object in comparison with the others, is connected with the position of 
         its nodes (see Fig. \ref{all}, panel F, middle column). For a minor body moving in a near-polar orbit, close encounters with major 
         planets are only possible in the vicinity of the nodes. When $\omega$ is close to 0\degr or 180\degr, the perihelion takes place 
         within the ecliptic plane with one node situated at aphelion and the other at perihelion; however, if $\omega$ is close to 90\degr 
         or 270\degr, the perihelion occurs well away from the plane. In the case of 2011~MM$_{4}$, when $\omega$ is 0\degr or 180\degr, 
         encounters with Saturn and Neptune take place at perihelion and aphelion; if $\omega$ is 90\degr or 270\degr, encounters with 
         Uranus are possible at both nodes in a manner similar to what is observed for 83982 Crantor (2002 GO$_{9}$), a well-studied Uranian 
         co-orbital (de la Fuente Marcos and de la Fuente Marcos 2013). The period of this oscillation of the nodes is nearly 700 kyr. 
         At present time, the closest approach to Saturn is at 1.08 AU but the Hill radius of Saturn is 0.41 AU; for Neptune, the closest 
         approach is at 0.94 AU but its Hill radius is 0.77 AU. This explains why its motion is rather stable, the planetary perturbations
         exerted on 2011~MM$_{4}$ are currently not strong enough to change the orbit appreciably.
%
%
%
%
%
%
%
%

      \subsection{2013~LU$_{28}$}    
         Asteroid 2013~LU$_{28}$ was discovered on 2013 June 8 by A. Boattini, R. E. Hill and J. A. Johnson observing with the 1.5-m 
         telescope of the Mt. Lemmon Survey (Bressi et al. 2013). It had a $V$ magnitude of 21.7 when first observed. Its orbit is very 
         poorly determined with just 27 observations acquired during 7 d and it is mainly included here to encourage follow-up observations.
         Its orbit is, in principle, the most eccentric of the ones discussed here. This object is also the largest, with $H$ = 8.1 mag or 
         50--150 km.
         
\begin{figure}
        \centering
         \includegraphics[width=\linewidth]{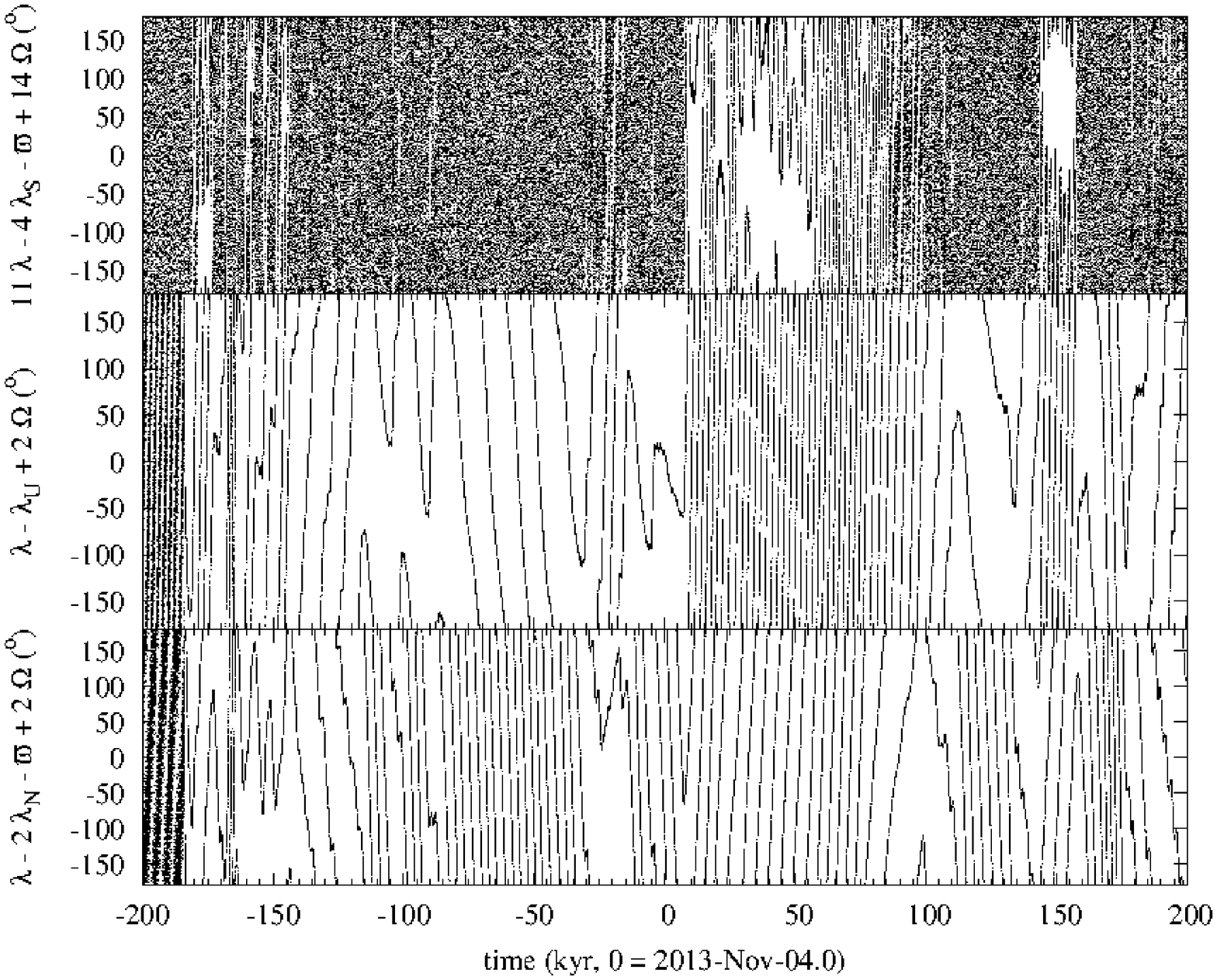}
         \caption{Evolution of resonant arguments for 2013~LU$_{28}$.
                 }
         \label{lu28}
      \end{figure}

         Due to the large uncertainty associated to the orbit of 2013~LU$_{28}$, our discussion here is more speculative. Assuming an object 
         moving in an orbit like the nominal one ($a$ = 19.25 AU, $e$ = 0.63, $i$ = 117\fdg00, $\Omega$ = 254\fdg57, $\omega$ = 179\fdg05 
         and $M$ = 297\fdg87), we observe that it is trapped in the 1:1 retrograde resonance with Uranus (asymmetric quasi-satellite 
         behaviour, see Fig. \ref{lu28}, middle panel). The object has remained in the neighbourhood of this resonance for nearly 200 kyr 
         but its dynamical status is rather unstable because it is submitted to additional resonances with Neptune and Saturn (see Fig. 
         \ref{lu28}, bottom and top panels), 2:1 and 4:11 retrograde, respectively. Our results are consistent with previous studies of the
         1:1 resonance at high inclination (see e.g. Brasser et al. 2004). All the control orbits suggest that it is a recent visitor from 
         the trans-Neptunian region. In this case, and given its very uncertain parameters, we have restricted our exploration of the 
         orbital domain to orbits very close to the nominal one. Its $e$-folding time is $\sim$3 kyr. 
%
%
%
%

 \section{Discussion}
    The distribution in absolute magnitude of known retrograde Centaurs with semi-major axis in the range 6--34 AU exhibits a clear deficit 
    of objects within 10--14 mag and a significant excess at $\sim$14 mag. This is consistent with theoretical expectations based on 
    collisional models (Fraser 2009). Objects with $H <$ 10 mag form a distinct group in dynamical terms. 

    The largest known retrograde Centaurs move in near-polar, comet-like orbits but no cometary activity has been detected yet on any of 
    them. The analysis of the short-term orbital evolution of these objects shows that the group is heterogeneous in origin. All of them are 
    currently trapped in transient retrograde resonances with the Jovian planets. Compared with the objects analysed by Morais and Namouni 
    (2013b), the duration of the resonant episodes of the objects studied here is longer as they are globally more stable. Asteroid (342842) 
    2008~YB$_{3}$ will escape towards the trans-Neptunian region in the near future and 2013~LU$_{28}$, a recent visitor from the 
    trans-Neptunian region, is the first known Uranian retrograde co-orbital candidate (a quasi-satellite) but further observations are 
    necessary to absolutely confirm that this object is indeed a retrograde companion to Uranus. 

    Another interesting feature of the dynamics discussed above is that high-order resonances seem to be preferred by these initial 
    conditions characterized by very high inclinations. This is, however, not surprising as pointed out by, for example, Robutel and Laskar 
    (2001) or Gallardo (2006). The strength of high-order resonances in general is greater for high-inclination orbits. At high inclination, 
    high-order resonances are strong enough to capture objects for relatively long periods of time. In our case, the high eccentricities are 
    also a contributing factor to produce stronger dynamical effects (see e.g. Gallardo 2006). Gallardo (2006) has found that, for a given 
    eccentricity and due to short-period planetary perturbations, high-inclination resonant orbits have a higher probability of survival 
    than low-inclination ones. This is also consistent with the comparatively weaker stability observed for the retrograde resonances 
    analysed by Morais and Namouni (2013b) and pointed out above; the orbits followed by the objects studied in their paper are not almost 
    perpendicular to the ecliptic as in our case and their resonances are low order. In general, chaotic diffusion is enhanced for 
    high-inclination, high-eccentricity orbits submitted to high-order resonances. 

    Asteroids 342842 and 2013~LU$_{28}$ are not long-term stable but 2011~MM$_{4}$ may be considerably more stable. In order to study the 
    long-term stability of 2011~MM$_{4}$ we use the Regularized Mixed Variable Symplectic (RMVS) integrator ({\it swift-rmvs3}) which is 
    part of the SWIFT package (Levison and Duncan 1994). The SWIFT package is available from H. Levison web 
    site.\footnote{http://www.boulder.swri.edu/$\sim$hal/swift.html} The physical model here is similar to the one described in Sect. 3 but 
    now the Earth and the Moon are replaced by the barycentre of the system. The calculations do not consider Mercury or any asteroids but 
    they include the barycentre of the Pluto-Charon system, use a timestep of 50 d and run for $\pm$5.0 Gyr. Relative errors in the total 
    energy are $< 4.0 \times 10^{-8}$. An ejection distance of 10000 AU from the Sun was used; any test particle that reached that distance 
    was removed from the calculations. As in the previous simulations, we have followed the evolution of 50 control orbits obtained from the 
    nominal orbital elements of 2011~MM$_{4}$ and their quoted uncertainties (see Table \ref{elements}). These orbital elements have been 
    sampled within 3$\sigma$ from a six-dimensional Gaussian distribution around the nominal ones. Results from RMVS and Hermite are 
    consistent within the applicable timeframe.

    Horner et al. (2004a, 2004b) studied the dynamical half-lives of prograde Centaurs and found values in the range 0.54--32 Myr. The mean 
    half-life of their sample of Centaurs was 2.7 Myr. For a given orbit, a large number of control orbits, $N_{o}$, with parameters close 
    to those of the nominal one were integrated. These orbits were followed until they resulted in a collision with the Sun or the ejection 
    of the test particle from the Solar System. The number of control orbits, $N$, that still remain within the boundaries of the Solar 
    System (in our case, within 10000 AU from the Sun) after a time $t$, is given by $N = N_{o}\ {\Large\rm e}^{-\lambda\ t}$, where 
    $\lambda = 0.693/T_{1/2}$, $\lambda$ is the decay constant and $T_{1/2}$ is the dynamical half-life of the objects. 

    Using {\it swift-rmvs3}, we have applied the same approach to 2011~MM$_{4}$ in an attempt to study its long-term stability. Asteroid 
    2011~MM$_{4}$ is clearly far more dynamically stable than most Centaurs; its mean dynamical half-life is 23.12 Myr. However, the range 
    in dynamical half-lives forward in time is 5.26--310.32 Myr and for the past injection into its present orbit is 3.56--1795.35 Myr. In 
    other words, this object probably arrived to the region of the giant planets nearly 23 Myr ago and certainly not earlier than nearly 2 
    Gyr ago. Therefore, it is not a primordial minor body that has remained around its present location since the formation of the Solar 
    System like Jovian or Martian Trojans. It will be ejected also within 23 Myr and not later than 310 Myr into the future.

\begin{figure}
       \centering
        \includegraphics[width=\linewidth]{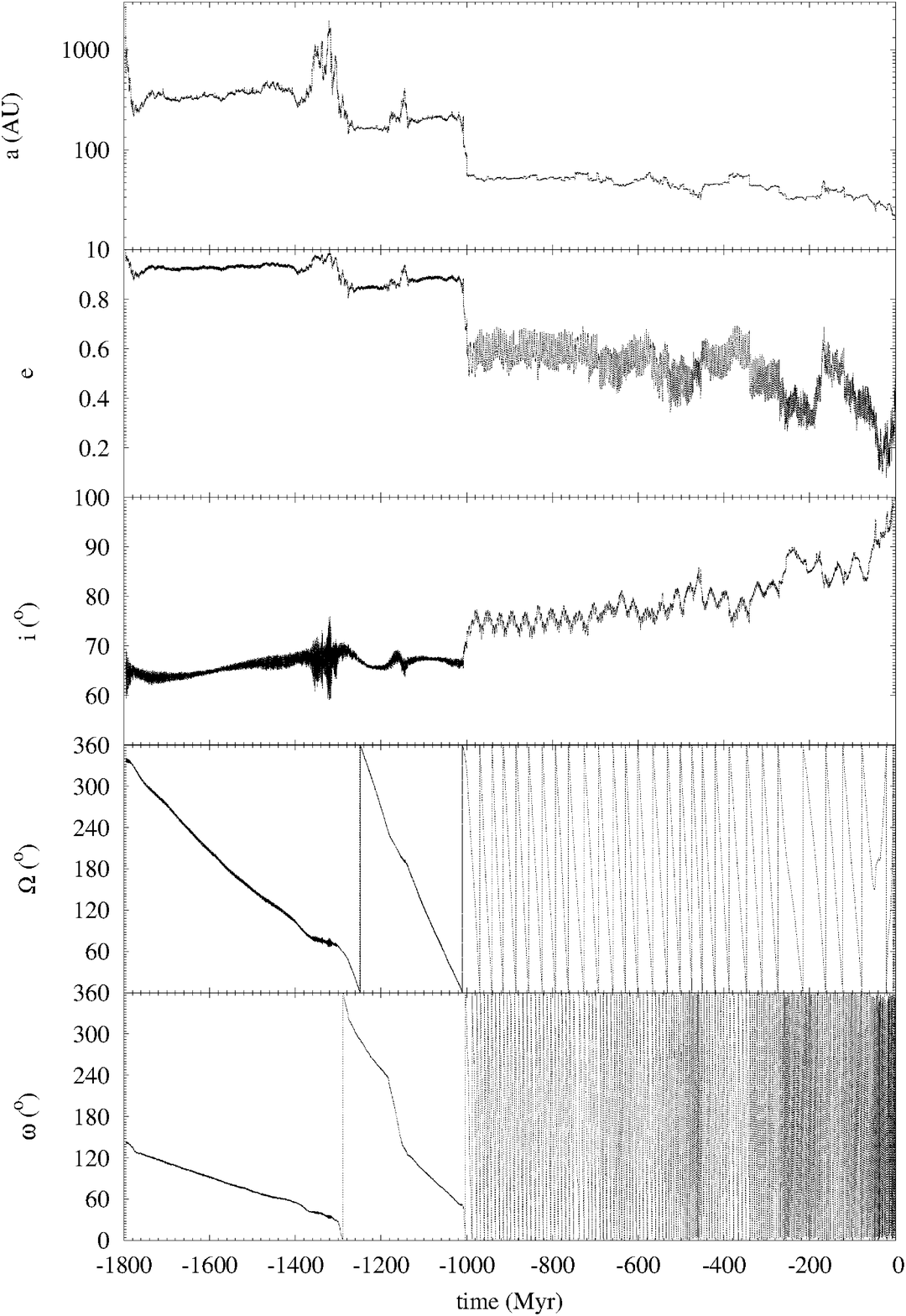}
        \caption{Time evolution of one of the control orbits of 2011~MM$_{4}$: semi-major axis, $a$, eccentricity, $e$, inclination, $i$, 
                 longitude of the perihelion, $\Omega$, and, argument of perihelion, $\omega$. The output time-step for this plot is 0.045 
                 Myr. These results correspond to the control orbit defined by $a$ = 21.1623 AU, $e$ = 0.47301, $i$ = 100.446\degr, $\Omega$ 
                 = 282.604\degr and $\omega$ = 7.224\degr.
                }
        \label{control}
     \end{figure}

    Among the control orbits, we have found one that is particularly interesting ($a$ = 21.1623 AU, $e$ = 0.47301, $i$ = 100.446\degr, 
    $\Omega$ = 282.604\degr and $\omega$ = 7.224\degr, see Fig. \ref{control}). It was very long lived into the past (1795.35 Myr) and its 
    associated dynamics may signal the presence of a reservoir of asteroids moving in high-inclination prograde orbits that are submitted to 
    a Kozai resonance as described by Gallardo (2006). In Fig. \ref{control}, the control orbit begins at a critical inclination of 
    $\sim$63\degr and very high eccentricity. From there, the eccentricity decreases and the initially prograde orbit turns into a 
    retrograde one. If, somehow, a reservoir of objects moving with orbital inclinations $\sim$63\degr and high eccentricities was created 
    (and may still survive) at 200--500 AU in the remote past, it may be a viable secondary source for retrograde Centaurs. 

    The dynamical evolutionary pathway uncovered by our calculations, even if it has a statistically low probability, shows that the Oort 
    cloud is not the only possible source region for retrograde Centaurs. The existence of additional stable reservoirs of rocky and icy 
    bodies at 100--300 AU cannot be simply casted aside. This interpretation is, however, tentative and should be treated 
    with caution, as there is recent circumstantial evidence on the existence of a massive planet at 250 AU from the Sun (Trujillo and 
    Sheppard 2014). If such planet does indeed exist, the architecture of the entire trans-Plutonian region could be very different from the 
    one assumed here, i.e. no truly massive bodies beyond Neptune. A massive perturber in the region 100--300 AU may affect the dynamical 
    evolution presented in Figure \ref{control} very significantly.

    Brasser et al. (2012), and Volk and Malhotra (2013) argued that retrograde or very high inclination Centaurs have their origin in the 
    Oort cloud. Fouchard et al. (2014) also pointed out that the Oort cloud, and in particular its inner part, is a potential source of high
    inclination Centaurs. Long-period comets are believed to have their origin in the Oort cloud (see e.g. Duncan 2008). Halley-type comets
    may come from the outer edge of the scattered disc (Levison et al. 2006) or the Oort cloud (Emel'yanenko et al. 2013). Jupiter-family 
    comets appear to have their origin in the scattered disk (Levison and Duncan 1997). For the group of minor bodies whose semi-major axes 
    are between 6 and 34 AU, the fraction of retrograde orbits is 11.3\%. There are 165 known comets with semi-major axes in the range 6--34 
    AU. Out of them, 20 objects or 12.1\% move in retrograde orbits. The fractions of retrograde orbits in both groups are more similar than 
    expected by chance alone. In principle, this fact argues strongly in favour of a dominant source in the Oort cloud for the retrograde 
    Centaurs but, again, the possible existence of trans-Plutonian planets may change the entire dynamical scenario.  
%
%
%
%

    Asteroid 2011~MM$_{4}$ is a promising candidate to be a representative of a remnant population of primordial planetesimals and it is a
    good candidate to be submitted to the nodal libration mechanism (Verrier and Evans 2008, 2009; Farago and Laskar 2010; Doolin and 
    Blundell 2011), at least temporarily. This mechanism, when active, can significantly enhance the stability of near-polar orbits. Minor 
    bodies like 2011~MM$_{4}$ may signal the size threshold for catastrophic collisions in the early Solar System. Returning to the issue of
    its origin, and as alternative to the Oort cloud or the hypothetical high-inclination reservoir pointed out above, we speculate that 
    this and other similar objects may be ancient ejected Uranian satellites (regular or irregular). The Uranian satellite system is the 
    least massive among those of the Jovian planets even if the largest one, Titania, is the eighth largest moon in the Solar System. Uranus 
    rotates nearly on its side with the axis lying nearly in its orbital plane. Uranus' large moons follow regular paths in equatorial-plane 
    orbits. Any object originally bound to Uranus and tracing an orbit lying nearly in its equatorial plane will follow an almost polar 
    orbit when ejected. 

 \section{Conclusions}
    In this research, we have used a comparative statistical analysis of the properties of known prograde and retrograde Centaurs to reveal 
    compelling reasons to single out the largest retrograde objects. The dynamical evolution of these objects has been subsequently studied
    using $N$-body simulations. The main conclusions of our work can be summarized as follows:
    \begin{itemize}
       \item The distribution in absolute magnitude of known retrograde Centaurs with semi-major axis in the range 6--34 AU exhibits a clear 
             deficit of objects in the range 10-14 mag and a significant excess at $\sim$14 mag. 
       \item The largest retrograde Centaurs move in almost polar, very eccentric paths with their nodal points currently located near 
             perihelion and aphelion. Within the retrograde Centaur population, they are clear outliers both in terms of dynamics and size.
       \item The largest retrograde Centaurs are currently trapped in transient retrograde, mostly high-order, resonances with the Jovian 
             planets. These resonances are strong enough to induce substantial chaotic diffusion and make their orbits dynamically unstable.
       \item Asteroid (342842) 2008~YB$_{3}$ appears to be significantly less dynamically stable than usually assumed.
       \item Asteroid 2011~MM$_{4}$ may be a visitor from the Oort cloud but an origin in a relatively stable closer reservoir cannot be 
             ruled out.
       \item Asteroid 2013~LU$_{28}$ is the first known Uranian retrograde co-orbital candidate (a quasi-satellite).
       \item This population may have its origin in the Oort cloud but alternative or concurrent sources could be ancient ejected Uranian 
             satellites or a hypothetical reservoir of minor bodies moving in high-inclination, high-eccentricity orbits at 200--500 AU and
             submitted to the Kozai resonance.
    \end{itemize}
    Our dynamical analysis has been restricted to just three objects and their associated orbital parameter space. However, a number of 
    distinctive properties that should be common to any retrograde Centaurs have been identified. In particular, the role of high-order
    resonances at high inclination as anticipated by Gallardo (2006) seems to be determinant on the short-term evolution of this population.

 \acknowledgments
    We would like to thank S.~J.~Aarseth, H.~F.~Levison and M.~J.~Duncan for providing the codes used in this research, and the referee for 
    his/her quick and helpful report. This work was partially supported by the Spanish `Comunidad de Madrid' under grant CAM S2009/ESP-1496. 
    We thank M. J. Fern\'andez-Figueroa, M. Rego Fern\'andez and the Department of Astrophysics of the Universidad Complutense de Madrid 
    (UCM) for providing computing facilities. Most of the calculations and part of the data analysis were completed on the `Servidor Central 
    de C\'alculo' of the UCM and we thank S. Cano Als\'ua for his help during this stage. In preparation of this paper, we made use of the 
    NASA Astrophysics Data System, the ASTRO-PH e-print server and the MPC data server.


\begin{thebibliography}{}
    \bibitem{AA03} Aarseth, S.J.:
            Gravitational N-body simulations.
            Cambridge University Press, Cambridge, p.\ 27 (2003)
    \bibitem{AL13} Alexandersen, M., Gladman, B., Greenstreet, S., Kavelaars, J.J., Petit, J.-M., Gwyn, S.:
            Science\ {\bf 341}, 994 (2013)
    \bibitem{BA13} Bauer, J.M., et al.:
            \apj\ {\bf 773}, 22 (2013)
    \bibitem{BE14} Belton, M.J.S.:
            \icarus\ {\bf 231}, 168 (2014)
    \bibitem{BE04} Bernstein, G.M., Trilling, D.E., Allen, R.L., Brown, M.E., Holman, M., Malhotra, R.:
            \aj\ {\bf 128}, 1364 (2004)
    \bibitem{BR04} Brasser, R., Heggie, D.C., Mikkola, S.:
             Celest. Mech. Dyn. Astron.\ {\bf 88}, 123 (2004)
    \bibitem{BR12} Brasser, R., Schwamb, M.E., Lykawka, P.S., Gomes, R.S.:
            \mnras\ {\bf 420}, 3396 (2012)
    \bibitem{BR13} Bressi, T.H., et al.:
            MPEC 2013-L69 (2013)
    \bibitem{CM07} Charnoz, S., Morbidelli, A.:
            \icarus\ {\bf 188}, 468 (2007)
    \bibitem{F12a} de la Fuente Marcos, C., de la Fuente Marcos, R.:
            \mnras\ {\bf 427}, 728 (2012a)
    \bibitem{F12b} de la Fuente Marcos, C., de la Fuente Marcos, R.:
            \mnras\ {\bf 427}, L85 (2012b)
    \bibitem{FF13} de la Fuente Marcos, C., de la Fuente Marcos, R.:
            \aap\ {\bf 551}, A114 (2013)
    \bibitem{FF14} de la Fuente Marcos, C., de la Fuente Marcos, R.:
            \mnras\ {\bf 441}, 2280 (2014)
    \bibitem{DB07} Di Sisto, R.P., Brunini, A.:
            \icarus\ {\bf 190}, 224 (2007)
    \bibitem{DI10} Di Sisto, R.P., Brunini, A., de El\'{\i}a, G.C.:
            \aap\ {\bf 519}, A112 (2010)
    \bibitem{DB11} Doolin, S., Blundell, K. M.:
            \mnras\ {\bf 418}, 2656 (2011)
    \bibitem{DU08} Duncan, M. J.:
            \ssr\ {\bf 138}, 109 (2008)
    \bibitem{EM13} Emel'yanenko, V.V., Asher, D.J., Bailey, M.E.:
            Earth, Moon, and Planets\ {\bf 110}, 105 (2013) 
    \bibitem{FL10} Farago, F., Laskar, J.:
            \mnras\ {\bf 401}, 1189 (2010)
    \bibitem{FO14} Fouchard, M., Rickman, H., Froeschl\'e, C., Valsecchi, G. B.:
            \icarus\ {\bf 231}, 99 (2014)
    \bibitem{FR09} Fraser, W.C.:
            \apj\ {\bf 706}, 119 (2009)
    \bibitem{FK09} Fraser, W.C., Kavelaars, J.J.:
            \aj\ {\bf 137}, 72 (2009)
    \bibitem{FR14} Fraser, W.C., Brown, M.E., Morbidelli, A., Parker, A., Batygin, K.:
             \apj\ {\bf 782}, 100 (2014)
    \bibitem{FH08} Fuentes, C.I., Holman, M.J.:
            \aj\ {\bf 136}, 83 (2008)
    \bibitem{GA06} Gallardo, T.:
            \icarus\ {\bf 181}, 205 (2006)
    \bibitem{GI96} Giorgini, J.D., et~al.:
            \baas\ {\bf 28}, 1158 (1996)
    \bibitem{GO11} Gomes, R.S.:
            \icarus\ {\bf 215}, 661 (2011)
    \bibitem{HL10} Horner, J., Lykawka, P.S.:
            \mnras\ {\bf 402}, 13 (2010)
    \bibitem{H04a} Horner, J., Evans, N.W., Bailey, M.E.:
            \mnras\ {\bf 354}, 798 (2004a)
    \bibitem{H04b} Horner, J., Evans, N.W., Bailey, M.E.:
            \mnras\ {\bf 355}, 321 (2004b)
    \bibitem{KO62} Kozai, Y.:
            \aj\ {\bf 67}, 591 (1962)
    \bibitem{LD94} Levison, H., Duncan, M.J.:
            \icarus\ {\bf 108}, 18 (1994)
    \bibitem{LD97} Levison, H.F., Duncan, M.J.:
            \icarus\ {\bf 127}, 13 (1997)
    \bibitem{LE06} Levison, H.F., Duncan, M.J., Dones, L., Gladman, B.J.:
            \icarus\ {\bf 184}, 619 (2006)
    \bibitem{MA91} Makino, J.:
            \apj\ {\bf 369}, 200 (1991)
    \bibitem{MC08} McNaught, R.H., et al.:
            MPEC 2008-Y38 (2008)
    \bibitem{M13a} Morais, M.H.M., Namouni, F.:
            Celest. Mech. Dyn. Astron.\ {\bf 117}, 405 (2013a)
    \bibitem{M13b} Morais, M.H.M., Namouni, F.:
            \mnras\ {\bf 436}, L30 (2013b)
    \bibitem{PI13} Pinilla-Alonso, N., et al.;
            \aap\ {\bf 550}, A13 (2013)
    \bibitem{RL01} Robutel, P., Laskar, J.:
            \icarus\ {\bf 152}, 4 (2001)
    \bibitem{SC13} Schlichting, H.E., Fuentes, C.I., Trilling, D.E.:
            \aj\ {\bf 146}, 36 (2013)
    \bibitem{SH13} Shankman, C., Gladman, B.J., Kaib, N., Kavelaars, J.J., Petit, J.M.:
            \apj\ {\bf 764}, L2 (2013)
    \bibitem{SH10} Sheppard, S.S.: 
            \aj\ {\bf 139}, 1394 (2010)
    \bibitem{ST98} Standish, E.M.: 
            JPL Planetary and Lunar Ephemerides, DE405/LE405,
            Interoffice Memo. 312.F-98-048, Jet Propulsion Laboratory, Pasadena, CA, USA (1998)
    \bibitem{TS14} Trujillo, C.A., Sheppard, S.S.:
            \nat\ {\bf 507}, 471 (2014)
    \bibitem{VE08} Verrier, P.E., Evans, N.W.:
            \mnras\ {\bf 390}, 1377 (2008)
    \bibitem{VE09} Verrier, P.E., Evans, N.W.: 
            \mnras\ {\bf 394}, 1721 (2009)
    \bibitem{VM13} Volk, K., Malhotra, R.:
            \icarus\ {\bf 224}, 66 (2013)
    \bibitem{WO11} Woodworth, D., et al.:
            MPEC 2013-N02 (2011)
 \end{thebibliography}
\end{document}